# INVESTIGATING THE MULTIDIMENSIONAL SEPARATION BEHAVIOR OF PARTICLES IN A CYCLOSIZER SETTING – A CASE STUDY ON CALCITE, FLUORITE AND MAGNESITE

*Johanna Sygusch[1], Thomas Wilhelm[2], Orkun Furat[2], Volker Schmidt[2], Martin Rudolph[1]*

*1 Helmholtz-Zentrum Dresden-Rossendorf, Helmholtz Institute Freiberg for Resource Technology, Chemnitzer Straße 40, 09599 Freiberg, Germany*

*2 Ulm University, Institute of Stochastics, Helmholtzstraße 18, 89069 Ulm, Germany*

**Abstract**

Particle separation is typically investigated regarding one particulate property only. Virtually all separation processes, however, act on various particle properties in different ways. Modern particle analytical modalities enable a statistically meaningful multidimensional particle characterization. Within this study, individual particle fractions of magnesite, calcite and fluorite (-71 µm) are processed via the turbulent cross-flow separator cascade Cyclosizer (M16, MARC Technologies Pty Ltd), consisting of 5 hydrocyclones, thus producing 5 different product streams. Particle characterization is achieved via dynamic image analysis from which information on the particle shape and size is obtained. Using this data, bivariate Tromp functions are computed, which show the combined effect of the particle descriptors of roundness and area-equivalent diameter on the separation behavior. While the first cyclones recover predominantly coarse particles with high roundness values, fine particles with varying roundness are recovered in the latter cyclones.

**INTRODUCTION**

In the minerals processing industry, hydrocyclones are a widely used tool with various applications within the processing stages. They are efficient for the classification of particles with sizes between 5 µm and 250 µm and are commonly placed in a milling circuit or for desliming of the flotation feed, especially for Kaolin processing, where a high amount of fines leads to several issues for the separation, e.g. slime coating (Trahar, 1981). Via the turbulent cross-flow hydroclassification, coarser particles tend to follow the outer streamlines and pass the underflow discharge, whereas finer particles tend to follow the inner upward streamline and pass the overflow discharge (Schubert, 1989). This effect is mainly influenced by the particle size and the particle density, as both properties affect the settling velocity of the particles, which determines to a large extent which streamlines the particles follow. The turbulent flow conditions within the hydrocyclone are influenced by its dimensions, such as diameter, cone angle or geometry of the underflow or the feed nozzle, but also by the pulp density of the feed suspension, as particles are experiencing hindered settling due to particle swam effects, thus creating challenges for the development of separation models (Schubert, 2010). Adding to this, is the complexity of the particle system, as particles usually vary in size, shape and composition. Typically, these challenges are addressed by model simplification or by including a mean parameter for describing the whole particle system (Schubert, 2003; Schubert, 2010).

In order to integrate the particle properties into the model, a better understanding of how certain particle properties influence the process is crucial. Kashiwaya et al. (2012) investigated the behavior of plate-like, block-shaped and spherical particles in a hydrocyclone as well as in a hydrocyclone cascade. They showed that the separation outcome is significantly influenced by the particle shape, as particles that were expected to be recovered in the underflow product on the basis of their size, were actually recovered in the overflow product, which they presumed to be a result of their plate-like shape. However, when it comes to the effect of particle shape on the particle behavior in a hydrocyclone there are very few studies found in literature (Endoh et al., 1994; Kashiwaya et al., 2012).

Additionally, since the particle separation behavior is typically governed by multiple properties rather than a single one, it is crucial to investigate their combined effect. Rather than using the commonly known univariate partition curves, also known as univariate Tromp functions, that display the separation function based on a single particle property, bivariate Tromp functions provide a more detailed insight into the influence of two particle descriptors at the same time and thus offer a more detailed look into their complex interactions. In order to compute bivariate Tromp functions, particle discrete information on the individual material streams is needed. One way of obtaining this information is via automated mineralogy (MLA) or dynamic image analysis. While MLA allows the analysis of different phases, thus being more suitable for complex mineral ores, dynamic image analysis provides information on the particle shape and size, but not on the particle composition and is therefore more suitable for single mineral analysis (Li & Iskander, 2020; Schulz et al., 2020).

Several studies have shown that multidimensional Tromp functions can be computed based on different approaches. Schach et al. (2019), for example, used kernel density estimators to determine bivariate Tromp functions for the separation in a Falcon separator according to particle size and mass density. Wilhelm et al. (2023), on the other hand, investigate the influence of particle shape and size for particles with different levels of wettability on their separation by flotation, using bivariate Tromp functions computed by using a parametric modeling approach.

Within this study, individual particle fractions of calcite, fluorite and magnesite are processed in a turbulent cross-flow separator cascade Cyclosizer, from which five different product fractions are obtained. Since the particle systems used for this study consist of one material only and are processed individually, dynamic image analysis is chosen as the characterization technique of the product fractions, as it is less time-consuming than automated mineralogy and still provides sufficient information on the relevant particle property information, here size and shape. Based on the particle discrete data obtained from image measurements, bivariate Tromp functions are computed regarding roundness and area-equivalent diameter. These functions are determined using kernel density estimators. In this way, the influence of particle size and shape on the particle separation behavior in a turbulent cross-flow cascade is investigated.

## MATERIAL AND METHODS

### Material

Calcite ($\rho = 2.71$ g/cm3), Fluorite ($\rho = 3.18$ g/cm3) and Magnesite ($\rho = 3.05$ g/cm3), all purchased from Krantz Rheinisches Mineralien-Kontor, Germany, are used as materials in this study. Suitable size fractions of -71 µm are obtained by dry sieving prior to the separation tests. Each particle system is represented by a set of particle descriptor vectors, where each vector contains a size and shape descriptor characterizing a single particle. The size of a particle is described by its area-equivalent diameter with respect to Eq.1, whereas the shape is described by its roundness with respect to Eq. 2. Note that for each particle both descriptors are computed from its projection obtained from dynamic image analysis. These sets of particle descriptor vectors are modeled by bivariate probability densities, which are displayed in Figure 1.

$$area - equivalent\ diameter\ d_A = 2\sqrt{\frac{projected\ area}{\pi}} \quad (1)$$

$$roundness\ r = \frac{4\pi \cdot area}{perimeter^2} \qquad (2)$$

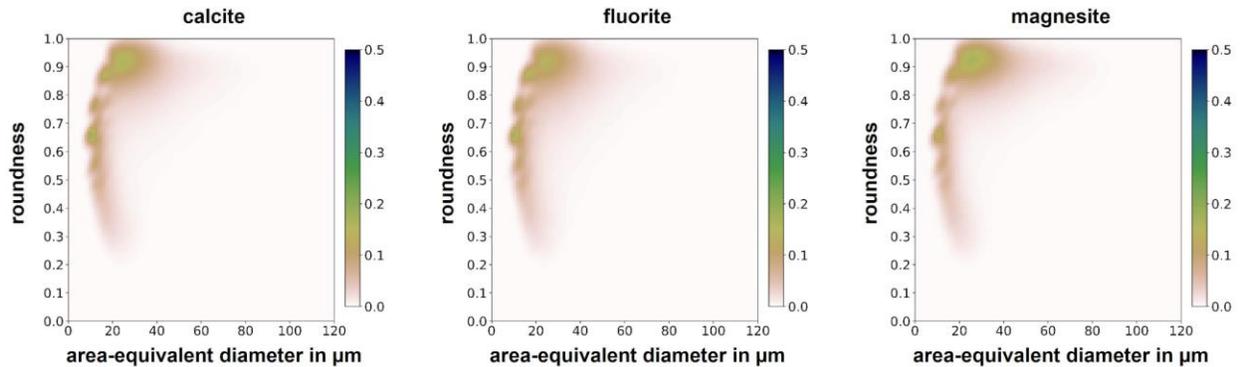

**Figure 1. Bivariate probability densities of roundness and area-equivalent diameter for calcite (left), fluorite (middle) and magnesite (right).**

**Turbulent cross-flow separator cascade**

The separation tests are carried out using a Cyclosizer M16 from MARC Technologies Pty Ltd, Australia, which is displayed in Figure 2. It consists of 5 inverse cyclones that are placed in a series, where the overflow of one cyclone will provide the feed for the next cyclone and the feed lines become successively smaller. Therefore, coarser particles tend to be trapped in the first cyclones, while finer particles are drawn along and tend to be retained in the later cyclones. The very fine particles that leave the last cyclone could have been collected with great effort. However, this was not the case for the presented study, resulting in the loss of these particles. The sample, around 25 g for each material, is dispersed in 150 ml of tap water for 10 min at 500 rpm on a stirring plate. It is then transferred into feed container and water is added until it is full. The shut feed container is inserted into the Cyclosizer set-up, the pump is turned on so that water is passing through all the cyclones and all remaining air bubbles are removed. The feed suspension is added to the stream over a period of 5 min with the control valve fully open. Afterwards, the elutriating time is set to 15 min with a flow of 11.6 l/min. After fully opening the control valve, the samples are taken from the cyclones, starting with the last cyclone no. 5. All suspensions are left to sediment, the supernatant is decanted and the remaining sample is dried in a drying cabinet at 70 °C for 24 hours and weighed in order to obtain the masses for each cyclone.

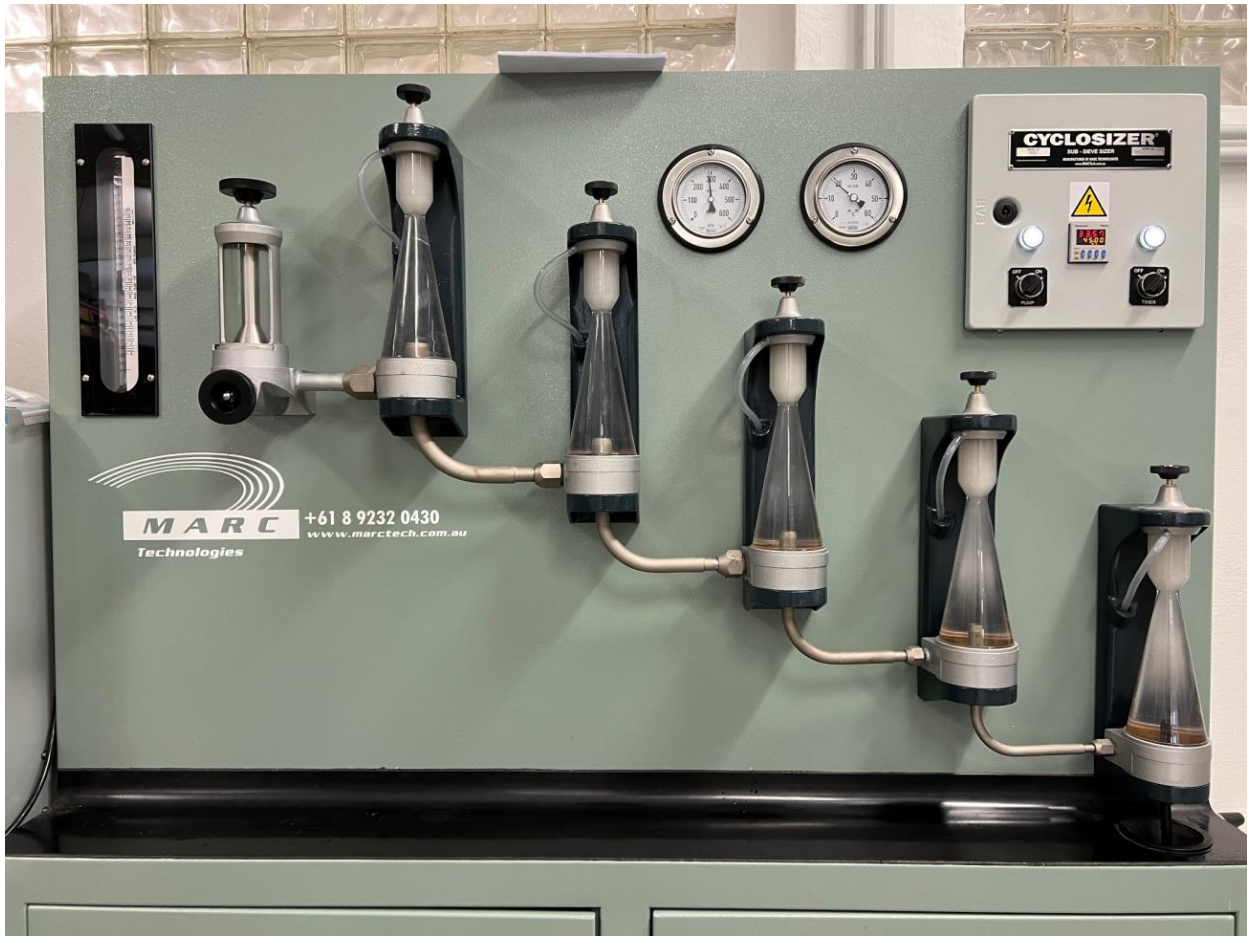

**Figure 2. The Cyclosizer M16 with the 5 cyclones used for the separation tests of this study.**

## Dynamic image analysis

Dynamic image analysis is carried out using the PICTOS from SYMPATEC, Germany. Videos of the measurements are exported, from which images are extracted using FFmpeg, an example is displayed in Fig. 3. The resulting images have a resolution of 4.972 µm per pixel. These images are then used to compute bivariate probability densities as well as bivariate Tromp functions in order to analyze the separation behavior of particles with respect to their size and shape. Characterization of the minerals from PICTOS image data, involves the application of image pre-processing steps. First, all particles with an area-equivalent diameter smaller than 2 pixels are excluded from further characterization. Subsequently, the presence of annulus-shaped particle projections, which exhibit a hole in the center is addressed. These projections are filled to ensure accurate computation of size and shape of the corresponding particle. However, some projections are shaped like an annulus segment, i.e. they are not fully closed. These projections are identified by computing their convexity factor and exclude all corresponding particles when the convexity factor is smaller than 0.3. A more detailed description of the convexity factor is presented by Chiu et al. (2013).

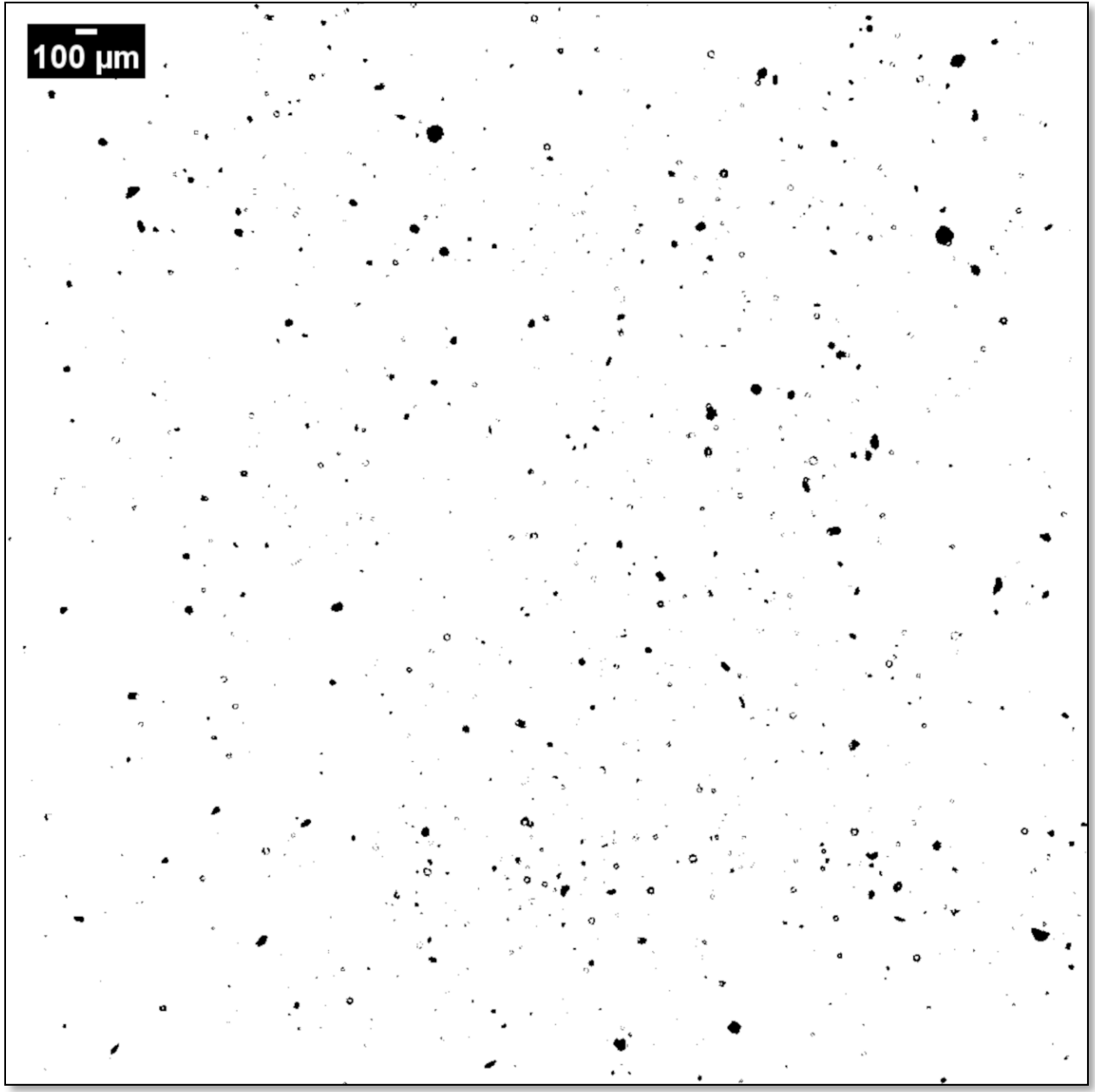

**Figure 3. Exemplary image of the calcite feed obtained with dynamic image analysis.**

## Computation of multivariate Tromp functions

In the present study, image data of the feed and of each of the five concentrates ($C_1$, $C_2$, $C_3$, $C_4$ and $C_5$) is available for each mineral (calcite, fluorite, magnesite). In order to evaluate the separation behavior of each particle system in the separation experiments, the entirety of particle descriptor vectors associated with particles in the feed material (F) is used to model the mass-weighted probability density $f_m^F$. Analogously, the probability densities $f_m^{C_i}$ are modeled for each concentrate $C_i$ with $i = 1, ... ,5$. Then, the mass-weighted probability density $f_m^C$ of particle descriptor vectors associated with particles in the entire concentrate (C) are computed as a convex combination of $f_m^{C_1}, ... , f_m^{C_5}$ as described in Wilhelm et al. (2023). This enables to compute the mass-weighted probability density $f_m^T$ of particle descriptor vectors associated with particles in the tailings (T) from the convex combination of $f_m^C$ and $f_m^T$ resulting in $f_m^F$ with mixing parameter $m_C/(m_C + m_T)$, where $m_C$ is the total mass of particles in (C) and $m_T$ the total mass of particles in the tailing T.

In the following, the separation of each cyclone is analyzed by means of Tromp functions. Therefore, exemplary for calcite, each cyclone is considered as an individual separation experiment with its individual feed, concentrate and tailings. More precisely, for each cyclone $i \in \{1,2,3,4,5\}$ probability densities $f_m^{F_i}$, $f_m^{C_i}$ and $f_m^{T_i}$ of particle descriptor vectors associated with particles in the feed, concentrate and tailings ($T_i$) of the $i$-th cyclone are determined, where the last cyclone is considered as first. For this cyclone ($i = 5$), the probability density $f_m^{T_5}$ equals to $f_m^T$ and $f_m^{C_5}$ has already been modeled from image data as described above. Then, the probability density $f_m^{F_5}$ is computed as a convex combination of $f_m^{T_5}$ and $f_m^{C_5}$ with the mixing parameter $m_{C_5}/(m_{C_5} + m_{T_5})$, where $m_{C_5}$ is the total mass of particles in the concentrate $C_5$ and $m_{T_5}$ the total mass of particles in the tailing $T_5$. For the next cyclone ($i = 4$), the probability density $f_m^{T_4}$ equals to $f_m^{F_5}$. Again, $f_m^{C_4}$ is already modeled as described above. Analogously, to the previous step $f_m^{F_4}$ is computed as a convex combination of $f_m^{T_4}$ and $f_m^{C_4}$. This procedure is repeated for each cyclone, where, for the first cyclone ($i = 1$), $f_m^{F_1}$ equals to $f_m^F$. This enables the computation of bivariate Tromp functions $T_i$ for each cyclone $i \in \{1,2,3,4,5\}$ as the ratio of $f_m^{C_i}$ and $f_m^{F_i}$ multiplied with the mass ratio of particles observed in each concentrate and feed. Then, the values $T_i(d_A, r)$ of the Tromp functions indicate the probability that a particle with descriptor vector $(d_A, r)$ is separated into the concentrate, as described in Wilhelm et al. (2023). With the same procedure we determine Tromp functions $T_i$ for each cyclone $i \in \{1,2,3,4,5\}$ for fluorite and magnesite.

In the present paper, $T_i$ is determined for all $i \in \{1,2,3,4,5\}$ and for each mineral. Furthermore, for visualization purposes all mass-weighted probability densities are transformed to number-weighted probability densities, as described in Wilhelm et al. (2023).

**RESULTS AND DISCUSSION**

The results of the individual mass fractions that are recovered in the different cyclones, plus the mass fraction of ultrafine particles that is lost, are depicted in Fig. 4. The distribution of the mass fractions is similar for all three minerals, with a decreasing trend from cyclone 1 to cyclone 5. The lost material, i.e. the particles that are too fine and pass the last cyclone, make up the largest mass fraction in all cases, with up to 41 % for magnesite. The $d_{50}$ values of particle size corresponding to the cyclones is displayed in Table 1, showing that very similar results are obtained for the three minerals. The $d_{50}$ values of the individual cyclones nicely show the sharp separation cut of the Cyclosizer, as defined particle size fractions are obtained that are only a few microns apart from each other.

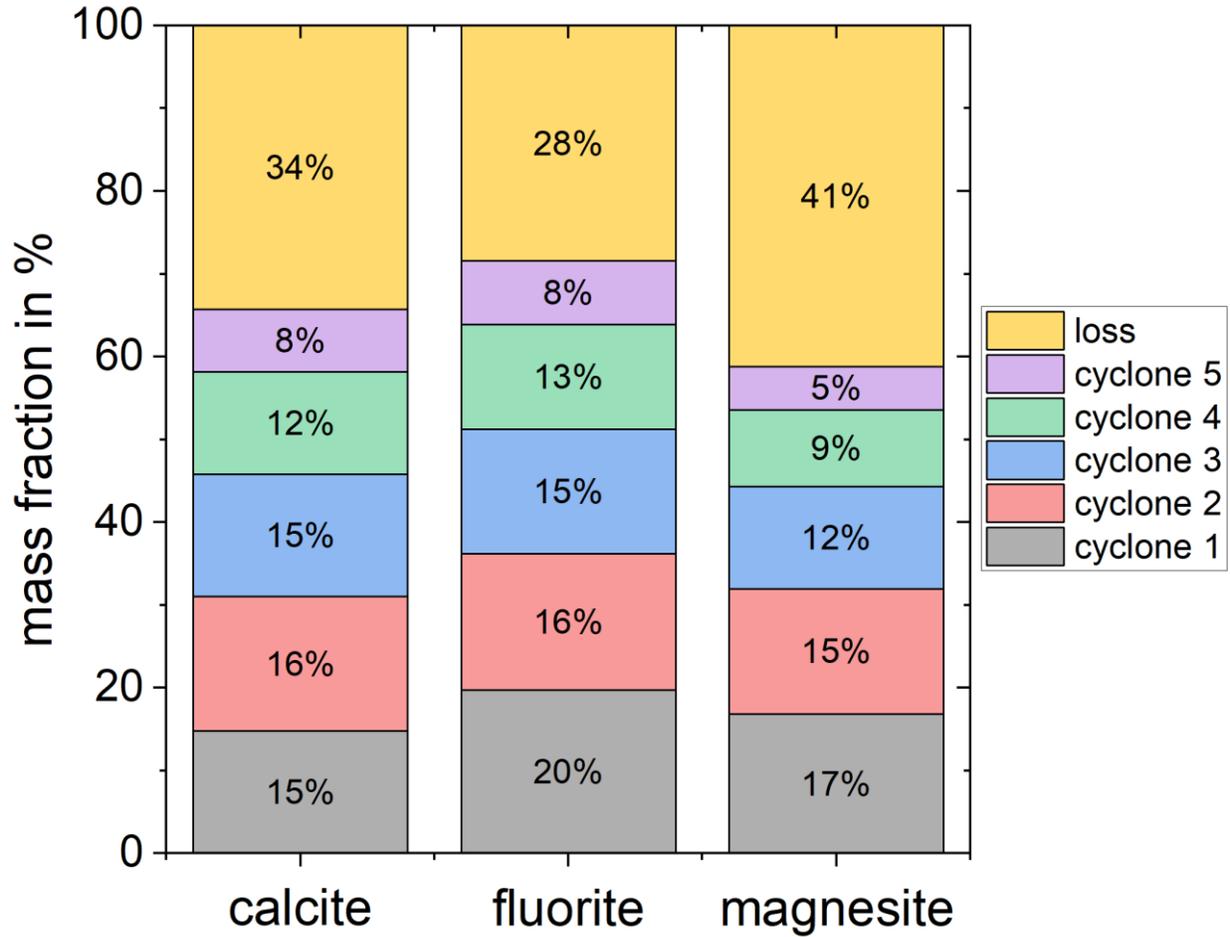

**Figure 4.** The mass fraction of the material for each cyclone, as well as the fraction that is lost for calcite, fluorite and magnesite.

**Table 1. The $d_{50}$ of each cyclone fraction for calcite, fluorite and magnesite obtained via laser diffraction**

| sample name | $d_{50}$ of cyclone no. in µm | | | | |
| --- | --- | --- | --- | --- | --- |
| | 1 | 2 | 3 | 4 | 5 |
| calcite | 62.47 | 49.57 | 36.34 | 25.14 | 17.92 |
| fluorite | 59.09 | 45.70 | 33.12 | 23.16 | 16.59 |
| magnesite | 59.38 | 45.31 | 32.78 | 22.00 | 15.79 |

So far, the classification for the different cyclones by size only has been reported. Figs. 5, 7 and 9 present the bivariate probability densities of particle shape, as roundness, and particle size, as area-equivalent diameter, for the feed and cyclone fractions of calcite, fluorite and magnesite, respectively. All three minerals have similar probability densities for their feed as well as the same trend for the individual cyclones. Particles in cyclone 1 are rather coarse and round with highest frequencies of particle sizes around 70 µm and roundness values around 0.9. Particles in cyclone 2 still show high values of roundness, but the dominant particle size is now in the range of 40 µm and 60 µm. This reduction in particle size is continued for cyclone 3, 4 and 5, with particle size fractions around 35 µm, 25 µm and 20 µm, respectively. However, while the particles in cyclone 3 mainly exhibit roundness values from around 0.8 to 1 with few exceptions, for all three minerals, an increasing number of particles with a wide range of roundness values down to 0.2 is observed for particles in cyclone 4 and cyclone 5. The bivariate probability densities show that the classification is significantly influenced by the particle size and shape, as coarser rounder particles tend to accumulate in the first cyclones, while finer particles with a wide range of roundness values are found in the latter cyclones, regardless of the mineral under investigation.

In order to investigate the combined influence of particle size and shape on the separation of the three minerals in a hydrocyclone cascade, bivariate Tromp functions have been computed. The Tromp functions for each individual cyclone for calcite, fluorite and magnesite are displayed in Figs. 6, 8 and 10, respectively. The color code gives information on whether a particle reports to the corresponding cyclone (high values of T) or if the particle reports to the fines, i.e. to the following cyclone or to the losses after cyclone 5 (low values of T). The Tromp functions for all three minerals show the same trend regarding particle size, as the maximum value of the Tromp functions shift from cyclone 1 to cyclone 5 down to finer particle sizes. The Tromp function for the first cyclone has high values for very coarse and very round particles, i.e. particles with these descriptor vectors are very likely to be recovered in the first cyclone, while it decreases significantly with decreasing particle size down to a value of zero for all minerals below 40 µm. At the same time, the value of the Tromp function also decreases with decreasing roundness, however this effect is not as pronounced as for particle size. The distribution of Tromp values for the second and third cyclone for calcite and magnesite exhibit similar values, except for particles with roundness around 0.95, where magnesite particles only have a probability of 60 %, whereas the calcite particles with this property have an almost 100 % probability of being recovered in the second cyclone. While this is also observed for very round fluorite particles, the probability declines with decreasing roundness, however it increases again for particles with very low roundness values of 0.1 – 0.4, which is not the case for calcite and magnesite. For the last two cyclones there is a significant shift observed for all three minerals, as the separation probability increases significantly for fine particles while it drops for coarse particles at the same time. Furthermore, particles over a wide range of roundness are recovered more preferably, as opposed to the first cyclones, where the Tromp function values were highest for particles with values close to a roundness of 1. This effect is most pronounced for the last cyclone, where separation probabilities of around 90 % are obtained for particles below 30 µm with roundness values of 0.1 – 1.

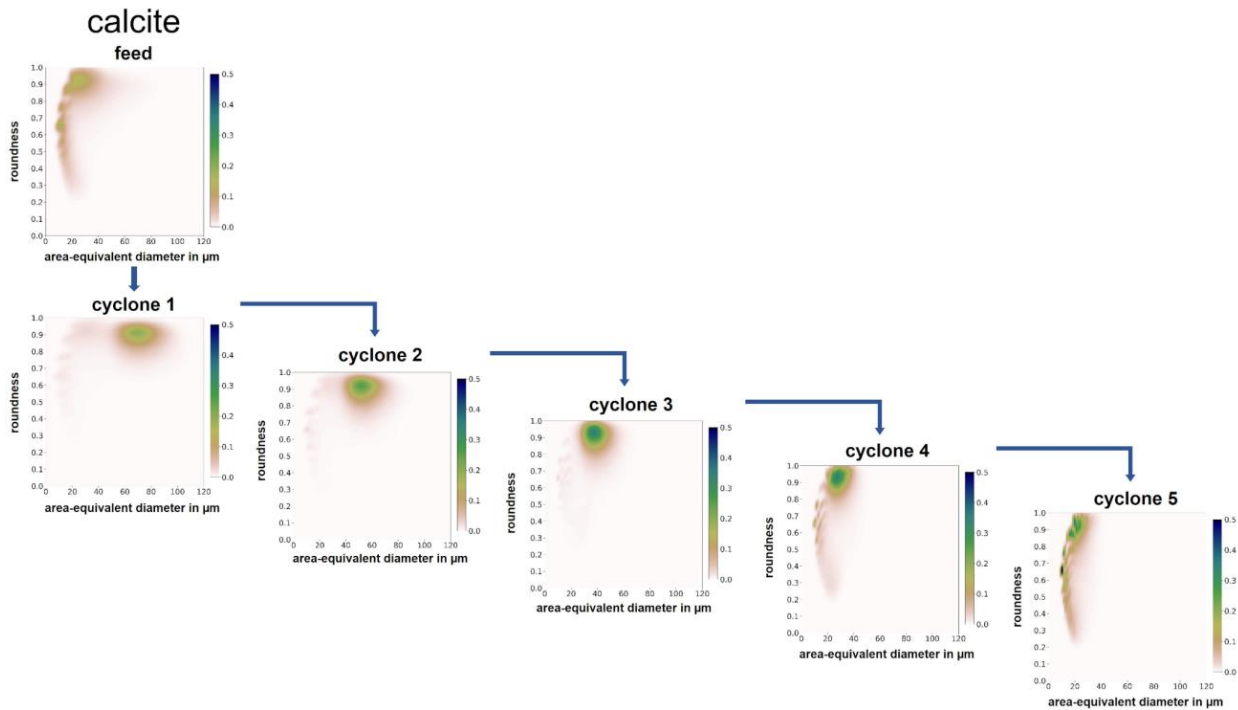

**Figure 5. Bivariate probability densities for particle roundness and area-equivalent diameter for the calcite feed and the individual cyclone fractions. The particle descriptors are obtained from dynamic image analysis.**

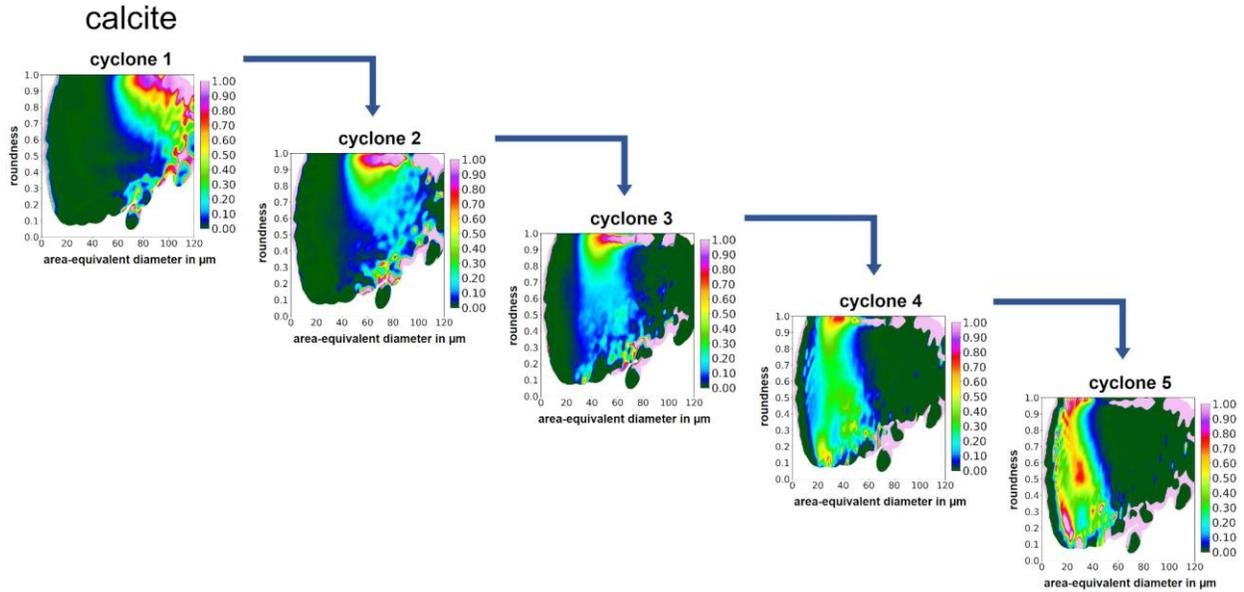

**Figure 6. Bivariate Tromp functions for the individual cyclones for separation tests using calcite. The color code indicates the value for the Tromp function T, i.e. the probability that the particle reports to the corresponding cyclone and has no units. The dark green color corresponds to a probability of T = 0.**

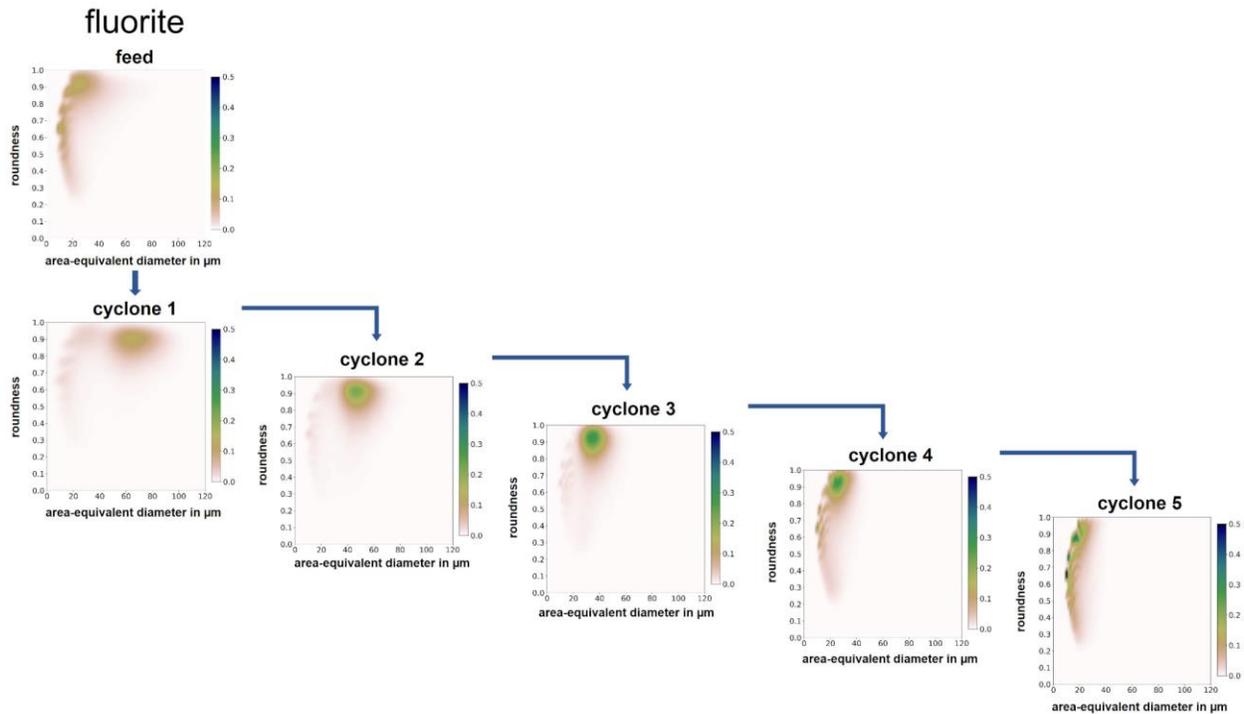

**Figure 7. Bivariate probability densities for particle roundness and area-equivalent diameter for the fluorite feed and the individual cyclone fractions. The particle descriptors are obtained from dynamic image analysis.**

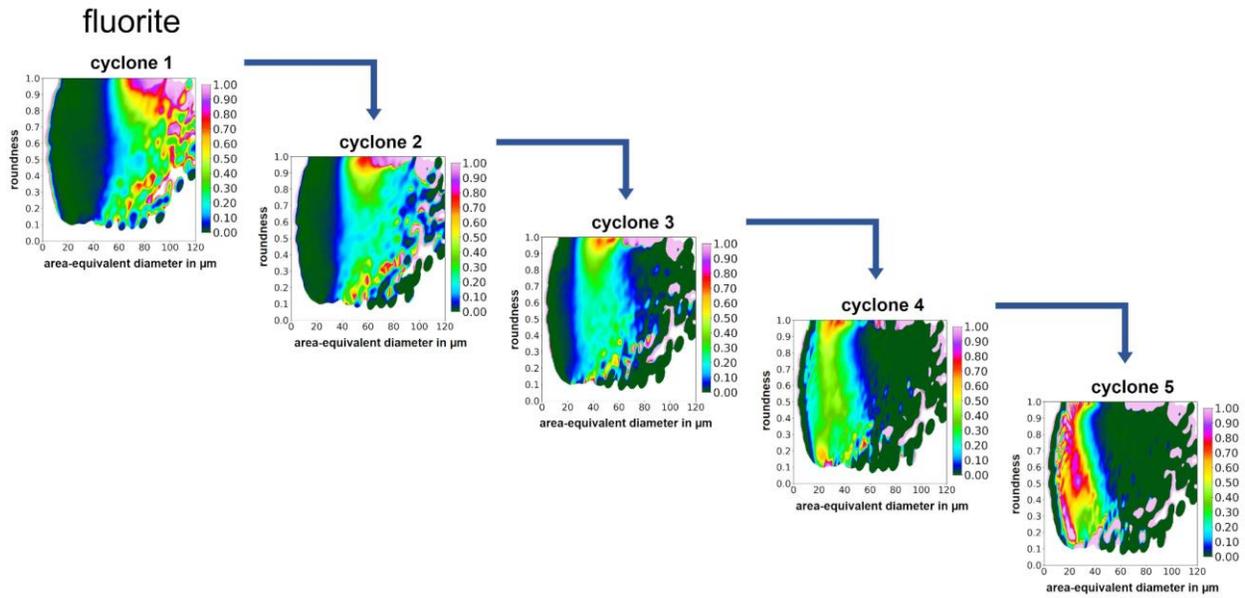

**Figure 8. Bivariate Tromp functions for the individual cyclones for separation tests using fluorite. The color code indicates the value for the Tromp function T, i.e. the probability that the particle reports to the corresponding cyclone and has no units. The dark green color corresponds to a probability of T = 0.**

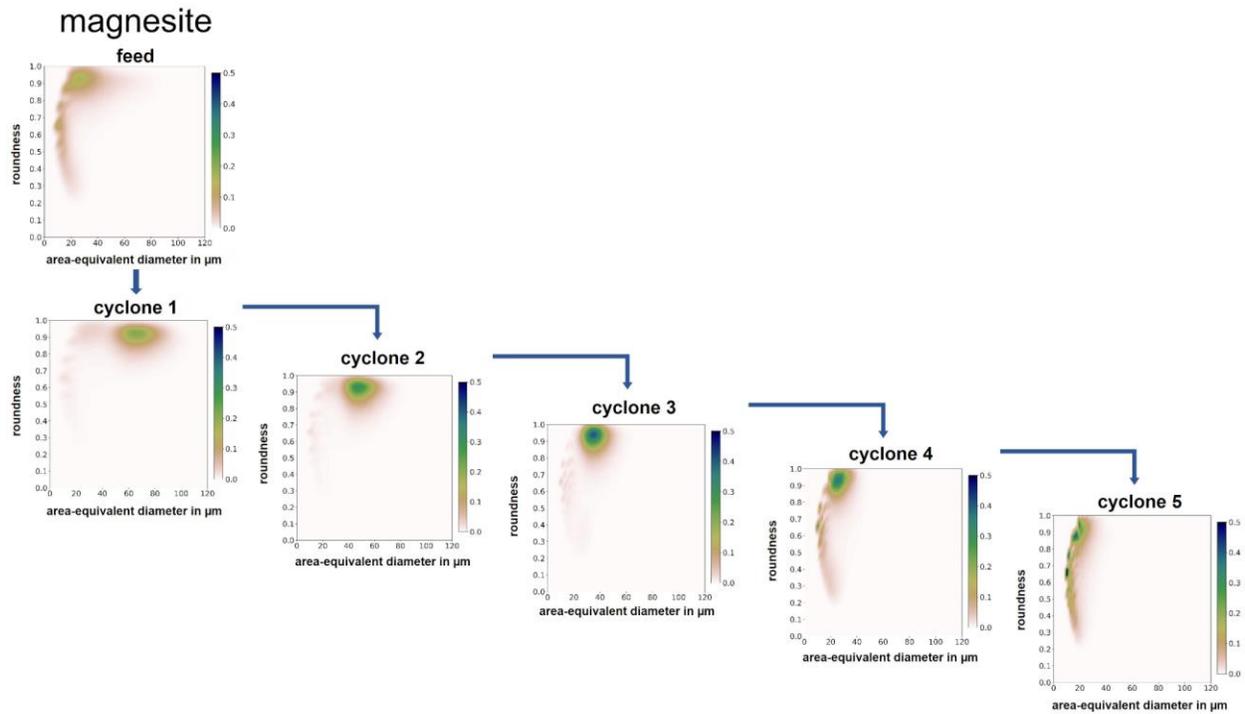

**Figure 9. Bivariate probability densities for particle roundness and area-equivalent diameter for the magnesite feed and the individual cyclone fractions. The particle descriptors are obtained from dynamic image analysis.**

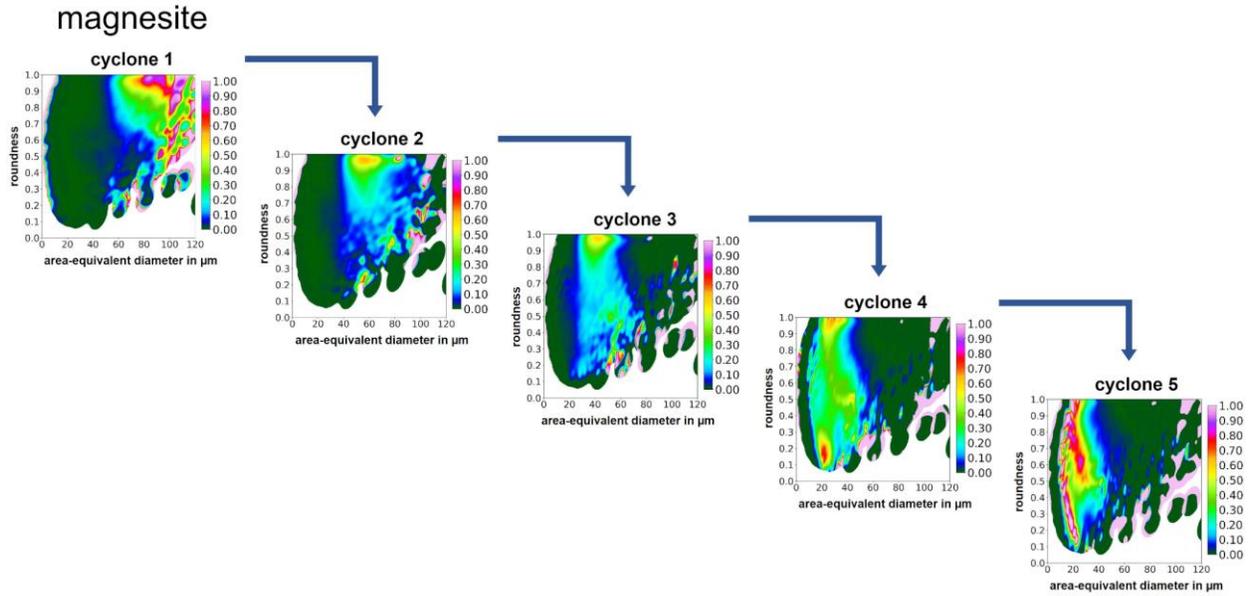

**Figure 10. Bivariate Tromp functions for the individual cyclones for separation tests using magnesite. The color code indicates the value for the Tromp function T, i.e. the probability that the particle reports to the corresponding cyclone and has no units. The dark green color corresponds to a probability of T = 0.**

## CONCLUSION

This study investigates the combined effect of particle size and shape, as area-equivalent diameter and roundness, respectively, on the separation of individual fractions of calcite, fluorite and magnesite in a hydrocyclone cascade. Five fractions are obtained after classification from the individual hydrocyclones, which are analyzed by laser diffraction for their particle size as well as dynamic image analysis, from which particle discrete information on particle size and shape is obtained. This particle discrete data is then used to compute bivariate Tromp functions with respect to particle descriptor vectors containing area-equivalent diameter and roundness. The results show that there is a significant influence of the particle size on the classification for all three minerals. From the first to the last cyclone, there is a clear shift in the particle size of the different particle fractions, from coarser particles being recovered in the first cyclone to finer particles in the last. The effect of roundness is not as pronounced as that of size, however, differences can be observed for the different particle fractions, as the first cyclones have a high separation probability for particles with very high roundness values, while fine particles over a large range of roundness report to the latter cyclones. Unfortunately, the three minerals investigated show a similar separation behavior, therefore, future studies should include particle systems that differ more significantly in shape, e.g. spheres or flakes, to reveal how these extreme variations of shape influence the separation. Furthermore, all particles that are too fine to be recovered in the last cyclone, which is actually the largest mass fraction here, are lost and so is the information on them. Hence, a possibility to recover that fraction needs to be found, in order to obtain information on all of the particles taking part in the process.

## FUNDING

This research is partially funded by the German Research Foundation (DFG) via the research projects RU 2184/1-2 within the priority program SPP 2045 "Highly specific and multidimensional fractionation of fine particle systems with technical relevance" and SCHM 997/45-1 within the priority program SPP 2315 "Engineered artificial minerals (EnAM) - A geo-metallurgical tool to recycle critical elements from waste streams".

**ACKNOWLEDGMENTS**

The authors would like to thank Rocco Naumann from the Helmholtz Institute Freiberg for Resource Technology for carrying out Cyclosizer tests as well as for analyzing the samples via laser diffraction and dynamic image analysis.